\documentclass[conference]{IEEEtran}
\IEEEoverridecommandlockouts
\usepackage{cite}
\usepackage{amsmath,amssymb,amsfonts}
\usepackage{algorithmic}
\usepackage{graphicx}
\usepackage{textcomp}
\usepackage{xcolor}
\usepackage{multirow}
\usepackage{array}
\usepackage{dblfloatfix}
\usepackage{soul}
\usepackage{array, makecell}
\usepackage{wrapfig}
\usepackage{xcolor}
\usepackage{hyperref}
\usepackage{multirow}
\usepackage{float}
\def\BibTeX{{\rm B\kern-.05em{\sc i\kern-.025em b}\kern-.08em
    T\kern-.1667em\lower.7ex\hbox{E}\kern-.125emX}}

\begin{document}

\title{Evading Deep Learning-Based Malware Detectors via Obfuscation: A Deep Reinforcement Learning Approach\\

\thanks{*: Corresponding author

Acknowledgments: This material is based upon work supported by the National Science Foundation (NSF) under the Secure and Trustworthy Cyberspace (1936370) program.}
}

\makeatletter

\newcommand{\linebreakand}{%
  \end{@IEEEauthorhalign}
  \hfill\mbox{}\par
  \par\hfill\begin{@IEEEauthorhalign}
}

\makeatother

\author{
\IEEEauthorblockN{Brian Etter*}
\IEEEauthorblockA{\textit{Department of Management Information Systems} \\
\textit{University of Arizona}\\
Tucson, USA \\
etterb@arizona.edu}

\and

\IEEEauthorblockN{James Lee Hu*}
\IEEEauthorblockA{\textit{Department of Management Information Systems} \\
\textit{University of Arizona}\\
Tucson, USA \\
jameshu@arizona.edu}

\linebreakand

\IEEEauthorblockN{Mohammadreza Ebrahimi}
\IEEEauthorblockA{\textit{School of Information Systems and Management} \\
\textit{University of South Florida}\\
Tampa, USA \\
ebrahimim@usf.edu}

\and

\IEEEauthorblockN{Weifeng Li}
\IEEEauthorblockA{\textit{Department of Management Information Systems} \\
\textit{University of Georgia}\\
Athens, USA \\
weifeng.li@uga.edu}

\linebreakand

\IEEEauthorblockN{Xin Li}
\IEEEauthorblockA{\textit{Department of Computer Science} \\
\textit{University of Arizona}\\
Tucson, USA \\
xinli2@arizona.edu}

\and

\IEEEauthorblockN{Hsinchun Chen}
\IEEEauthorblockA{\textit{Department of Management Information Systems} \\
\textit{University of Arizona}\\
Tucson, USA \\
hsinchun@arizona.edu}
}

\maketitle
\thispagestyle{plain}
\pagestyle{plain}

\begin{abstract}
Adversarial Malware Generation (AMG), the generation of adversarial malware variants to strengthen Deep Learning (DL)-based malware detectors has emerged as a crucial tool in the development of proactive cyberdefense. However, the majority of extant works offer subtle perturbations or additions to executable files and do not explore full-file obfuscation. In this study, we show that an open-source encryption tool coupled with a Reinforcement Learning (RL) framework can successfully obfuscate malware to evade state-of-the-art malware detection engines and outperform techniques that use advanced modification methods. Our results show that the proposed method improves the evasion rate from 27\%-49\% compared to widely-used state-of-the-art reinforcement learning-based methods.

\end{abstract}


\begin{table*}[t!]
\centering
\renewcommand{\arraystretch}{1.2}
\setlength{\abovecaptionskip}{0pt}
\setlength{\belowcaptionskip}{-10mm}
\caption{Selected Recent Significant AMG Studies}
\label{lit_overview}
\begin{tabular}{
|c<{\centering}
|c<{\centering}
|c<{\centering}
|c<{\centering}
|c<{\centering}
|c<{\centering}|}

\hline

\textbf{Year}&\textbf{Author(s)}&\textbf{Detector Type}&\textbf{Focus}&\textbf{Attack Method}&\textbf{Modification}\\

\hline

2022 & Song et al. \cite{song2022a} & DL-based & Attack & Reinforcement Learning & Additive\\

\hline

2021 & Javaheri et al. \cite{javaheri2021a} & Signature-based & Defense & Genetic Algorithm & Editing\\

\hline

2021 & Ebrahimi et al. \cite{ebrahimi2021a} & Signature-based & Attack & Deep Language Modeling & Editing\\

\hline

2020 & Demetrio et al. \cite{demetrio2020a} & DL-based & Attack & Genetic Algorithm & Additive\\

\hline

2019 & Park et al. \cite{park2019a} & DL-based & Attack & Dynamic Programming & Additive\\

\hline

2019 & Castro et al. \cite{castro2019a} & Signature-based & Attack & Random Perturbations & Additive\\

\hline

2019 & Suciu et al. \cite{suciu2019a} & DL-based & Attack & Benign Sequence Append & Additive\\

\hline

2019 & Rosenberg et al. \cite{rosenberg2019a} & DL-based & Defense & GAN & Additive\\

\hline

2018 & Anderson et al. \cite{anderson2018a} & Signature-based & Attack & Deep Reinforcement Learning & Editing\\

\hline

2018 & Dey et al. \cite{dey2018a} & Signature-based & Attack & Genetic Programming & Additive\\

\hline

\end{tabular}
\vspace{2mm}

{\centering \textbf{Note:} GAN: Generative Adversarial Network\par}

\vspace{-5mm}
\end{table*}

\section{Introduction}

Detection and identification of malware is one of the top priorities in cybersecurity. However, as malware authors write more sophisticated malware, existing malware detectors become less efficient \cite{raff2018a}. At a high level, malware detectors typically fall into one of two categories: traditional signature-based detectors and, more recently, detectors incorporating Deep Learning (DL) architectures \cite{raff2018a}, \cite{suciu2019a}. While showing promise in the early detection of new malware variants, DL-based malware detectors are vulnerable to adversarial attacks, which involve the deliberate insertion of benign code alongside the malicious code in order to fool the detector into misclassifying the malware as a benign file, thus evading detection \cite{anderson2018a}. It has been shown that adversarially generated malware samples that successfully evade detection can be leveraged to retrain and enhance, or robustify, the detector, improving subsequent performance against future attacks \cite{ebrahimi2020a}. To this end, Adversarial Malware Generation (AMG) aims to robustify malware detectors through automated generation of, and training on, crafted adversarial malware variants \cite{ebrahimi2021a}.

The majority of current AMG studies focus on small additive or editing actions with very few exploring full-file obfuscation \cite{ling2021a}. We expect that these small perturbations ultimately decrease the covert nature of adversarial variants, with the full potential of AMG unrealized. Thus, changing the makeup of the entire malware file through full-file obfuscation has the potential to contribute to the AMG field. Additionally, this would better reflect the practices of real-world hackers, who often obfuscate an entire file through the use of powerful editing actions \cite{nachreiner2017a}, \cite{baltazar2022a}.

Compression and encryption are techniques used by real-world hackers and can be used in generating evasive variants \cite{badhwar2021a}. A multitude of tools and methods can apply such actions in the obfuscation of malware \cite{javaheri2021a}, generating a large action space. However, considering all possible combinations of these tools, methods, and their parameters leads to a larger combinatorial action space. It is worth noting that not all modifications necessarily lead to an increase in evasive capabilities \cite{ling2021a}. To address this, Reinforcement Learning (RL) provides a promising framework to search for effective action sequences, or strings of modifying actions, that result in evasive malware variants \cite{fang2019a}. In this study, we aim to present a deep RL-based framework coupled with prevailing open-source obfuscation tools for conducting effective adversarial malware generation.

The rest of this paper is organized as follows. First, we conduct a literature review of adversarial malware generation, obfuscation, and reinforcement learning. We then present the details of the components of the proposed framework. Next, we compare the proposed method against extant state-of-the-art AMG methods. Lastly, we highlight several promising future directions.

\section{Literature Review}
In this work, we review four areas of the literature. First, the AMG landscape is surveyed and used as a foundation for the study. Second, we explore obfuscation methodologies and tools with an emphasis on open-source availability and capacity to automate. Third, RL is reviewed as a means to automate the generation of adversarial samples and identify evasive variants. Fourth, we explore Deep Q-Networks (DQN) and their role in model effectiveness and efficiency.
\subsection{Adversarial Malware Generation}

Malware detectors, by and large, fall into one of two categories, signature-based or DL-based. Signature-based detectors utilize libraries containing code known to be malicious and look for these sequences of code in the target file \cite{anderson2018a}, \cite{ebrahimi2021a}, \cite{javaheri2021a}, \cite{castro2019a}, \cite{dey2018a}; whereas DL-based detectors detect malicious files based on a learned representation of bytes\color{black} \cite{raff2018a}, \cite{suciu2019a}, \cite{song2022a}, \cite{demetrio2020a}, \cite{park2019a}, \cite{rosenberg2019a}. Despite their ability in identifying new variants, DL-based detectors tend to be more vulnerable to adversarial variants.

Adversarial Malware Generation (AMG) is the practice of adversarially modifying malware samples with the goal of being misclassified as benign by DL-based malware detectors. The ultimate goal of AMG studies is to strengthen detectors by using the adversarially crafted samples that evade detection to retrain the model, thus robustifying the detector against future adversarial attacks \cite{anderson2018a}. We summarize several significant recent works based on the detector type, focus (attack or defense), attack method, and modification strategy in Table \ref{lit_overview}.

The AMG methods generate adversarial variants through modifying the original malware files. Such adversarial modifications fall into two categories of actions: additive or editing. Additive actions are predominantly the appending of benign bytes, or sequences, to a malicious file \cite{suciu2019a}, \cite{song2022a}, \cite{dey2018a}, \cite{rosenberg2019a}. Random perturbations \cite{castro2019a}, insertion of benign bytes \cite{demetrio2020a}, \cite{dey2018a}, and insertion of dummy code \cite{park2019a} are also common; the latter tend to utilize more advanced DL-based models such as Generative Adversarial Networks (GAN), or Genetic or Dynamic Programming to facilitate the injection of benign bytes \cite{javaheri2021a}, \cite{demetrio2020a}, \cite{park2019a}, \cite{dey2018a}. The editing actions utilized rarely encompass more than a few sections of the malware file, and those that do are limited to file compression, or packing, through UPX \cite{anderson2018a}, \cite{ebrahimi2021a}. Due to the prevalence of UPX’s use in malware obfuscation however, most detectors see the act of compression/packing by UPX as malicious without scrutinizing the actual file, resulting in higher occurrences of false positives \cite{demetrio2020a}, \cite{aghakhani2020a}.

The key observations from Table \ref{lit_overview} are that there is a prevalence of additive actions \cite{suciu2019a}, \cite{song2022a}-\cite{rosenberg2019a} and a sparsity of editing actions \cite{anderson2018a}, \cite{ebrahimi2021a}. In particular, the additive actions are often applied using methods such as Genetic Programming (GP), Dynamic Programming (DP), Generative Adversarial Networks (GANs), etc., and may only be possible to implement by seasoned hackers or software engineers \cite{javaheri2021a}, \cite{demetrio2020a}, \cite{park2019a}, \cite{dey2018a}, \cite{rosenberg2019a}. In these studies, the majority of actions are designed to only modify small parts of the file or append a small number of bytes to avoid breaking its functionality, leaving the majority of the file’s makeup unchanged. However, hackers tend to use obfuscation techniques such as encryption, packing, or encoding to conceal the entirety of a file \cite{malviya2021a}. The use of open-source software to obfuscate the entire malware file may help to advance AMG techniques by leveraging readily available tools used in real-world scenarios. Thus, the inclusion of this approach may afford enhanced robustification of DL-based detectors \cite{mitre2021a}.

\subsection{Obfuscation}

\begin{table*}[ht!]
\centering
\renewcommand{\arraystretch}{1.2}
\setlength{\abovecaptionskip}{0pt}
\setlength{\belowcaptionskip}{-10mm}
\caption{Selected Major Obfuscation Tools from Literature}
\label{obfuscationTools}

\begin{tabular}{
|>{\centering\arraybackslash}c
|>{\centering\arraybackslash}c
|>{\centering\arraybackslash}c
|>{\centering\arraybackslash}c
|>{\centering\arraybackslash}c
|}

\hline
\textbf{Reference}                  & \textbf{OS / Commercial}    & \textbf{API / CLI}  & \textbf{Modification} & \textbf{Tool}               \\ \hline
\multirow{3}{*}{Aghakhani et al. \cite{aghakhani2020a}}   & \multirow{2}{*}{Commercial} & \multirow{2}{*}{No} & Compression           & Obsidium, PECompact, PELock \\ \cline{4-5} 
                                    &                             &                     & Encryption            & Themida                     \\ \cline{2-5} 
                                    & Open-Source                 & Yes                 & Compression           & Petite, UPX                 \\ \hline
Anderson et al. \cite{anderson2018a}                     & Open-Source                 & Yes                 & Compression           & UPX                         \\ \hline
\multirow{3}{*}{Bergenholtz et al. \cite{bergenholtz2020a}} & \multirow{2}{*}{Commercial} & \multirow{2}{*}{No} & Compression           & Obsidium, PECompact, PELock \\ \cline{4-5} 
                                    &                             &                     & Encryption            & Themida                     \\ \cline{2-5} 
                                    & Open-Source                 & Yes                 & Compression           & Petite, UPX                 \\ \hline
Halls \cite{halls2021a}                               & Open-Source                 & Yes                 & Encryption            & Darkarmour                  \\ \hline
Nichol \cite{unixpickle2016}                              & Open-Source                 & Yes                 & Encryption            & Gobfuscate                  \\ \hline
Metasploit Project    \cite{metasploit2015}              & Open-Source                 & Yes                 & Multiple              & MSFVenom                    \\ \hline

\end{tabular}
\vspace{-5mm}
\end{table*}

The functionality of a portable executable (PE) file is extremely sensitive to changes in the code, with the change of one byte being enough to render the malware file corrupt \cite{song2022a}. Because of this, the majority of editing actions, with the exception of compression by UPX, are designed to preserve the overall makeup of the malware file \cite{anderson2018a}. However, given the prevalence of malware file encryption in the real-world \cite{nachreiner2017a}, \cite{baltazar2022a}, we expect that adding this approach will extend the potential of AMG research to better emulate real-world scenarios. Thus, through the inclusion of readily available open-source tools that are used in real-world scenarios, DL-based detectors may experience heightened levels of robustification and be less susceptible to adversarial attacks \cite{mitre2021a}. Additionally, this approach would better reflect the actions of real-world hackers \cite{nachreiner2017a}, \cite{baltazar2022a}.
Tools and methods found in the current research encompass a small portion of the spectrum for obfuscation tools \cite{anderson2018a}, \cite{aghakhani2020a}, \cite{bergenholtz2020a}. Resources and subject matter experts (SMEs) outside of academic research also provide valuable insights about such commonly used tools and methods \cite{nachreiner2017a}, \cite{baltazar2022a}, \cite{mitre2021a}.


The literature largely documents tools that fall into two categories, open-source and commercial, with open-source tools typically found on GitHub, and commercial tools obtained directly from the developer. The ability to automate the obfuscation process through scripting and the command line interface (CLI) requires access to an application programming interface (API). Most tools found in extant literature utilize compression, encryption, or a combination of these two methods to obfuscate malware. Table \ref{obfuscationTools} summarizes major open-source and commercial grade tools used for obfuscation in literature or by the hacker community.

Three key points can be drawn from Table \ref{obfuscationTools}. First, within the commercial tools their API is not readily available with the main functionality locked behind a paywall or graphical user interface (GUI). The API call is of particular importance for automation purposes (interaction with the RL agent). Second, extant AMG research primarily documents tools utilized by the academic and/or research communities; with very few references to tools used by the hacker community outside of UPX (packing/compression) \cite{raff2018a}, \cite{anderson2018a}. There are several non-AMG studies done on packing tools and their ability to increase evasion of malware detectors ranging from commercial grade (paid or freemium) to open-source\cite{aghakhani2020a}, \cite{bergenholtz2020a}. However, due to the prevalence of packing in malware obfuscation, as noted previously, most detectors see the act of compression/packing, such as the use of the UPX packing tool, as malicious without scrutinizing the actual file. This is because UPX leaves specific artifacts in the code of a compressed PE file, which typically cause it to be flagged as potentially malicious \cite{demetrio2020a}, \cite{aghakhani2020a}, resulting in higher occurrences of false positives.

\begin{figure}[!h]
\centering
    \vspace{-3pt}
        \includegraphics[width=0.48\textwidth]{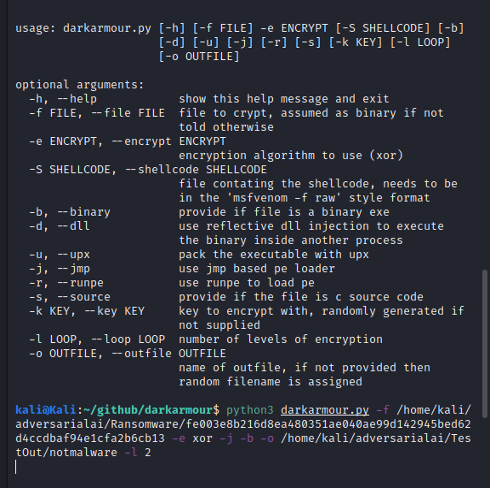}
    \vspace{-9pt}
    \caption{Interface of Darkarmour}
    \label{darkarmour}
\vspace{-5pt}
\end{figure}

Lastly, in our research, we considered recommendations from a Red Team SME who endorsed three tools: Darkarmour \cite{halls2021a}, Gobfuscate \cite{unixpickle2016}, and MSFVenom \cite{metasploit2015}. After a thorough evaluation, we opted for Darkarmour as the most suitable choice. This decision is underpinned by Darkarmour's simplicity of implementation, open-source accessibility, and availability through the CLI with an associated API. While Gobfuscate and MSFVenom received commendation from the SME, practical challenges, such as configuration and environment issues with Gobfuscate and misalignment with our research requirements in the case of MSFVenom as well as , led us to select Darkarmour as our preferred tool. Additionally, the Petite tool was excluded from consideration due to its absence of recent updates or active development, which did not align with our criteria. Accessing a tool's API via the CLI is crucial for automating and seamlessly integrating it into various workflows. This capability is demonstrated in Figure \ref{darkarmour} through Darkarmour's CLI, featuring a rich set of parameters suitable for automation. With a large number of available tools and parameters, RL has emerged as an ideal medium for navigating this action space and identifying optimal combinations of actions in the generation of evasive variants.

\subsection{Deep Reinforcement Learning}
Reinforcement Learning (RL) provides a useful framework to conduct goal-directed learning or optimization \cite{sutton2018a}. The goal of RL is to find the best action sequence, from a given set of actions, for a given input and for the agent to learn a (near) optimal policy; in this case the best set of modifications (actions) for a given malware file \cite{dey2018a}.

\begin{figure}[!h]
\centering
    \vspace{-9pt}
        \includegraphics[width=0.45\textwidth]{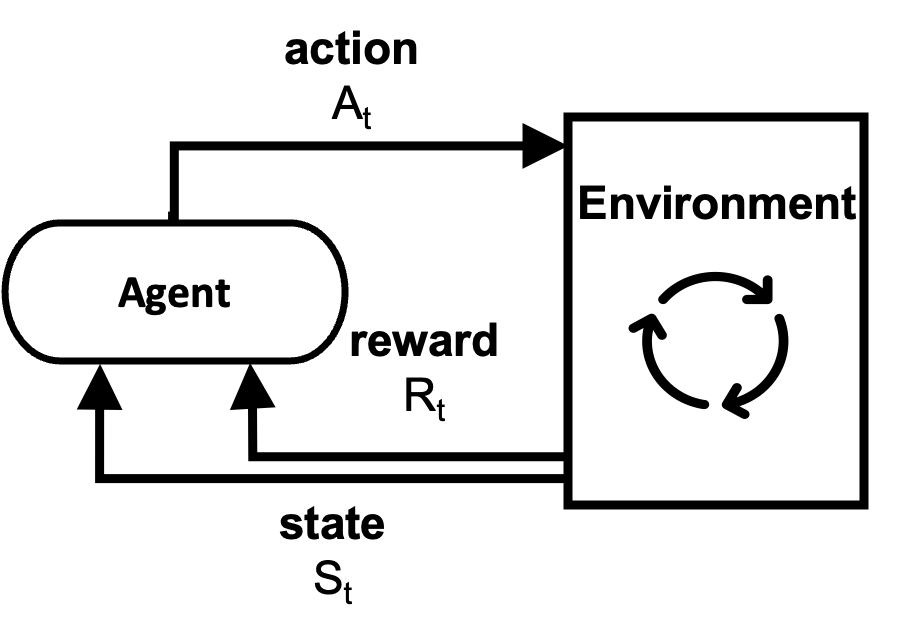}
    \vspace{-9pt}
    \caption{RL agent-environment interaction}
    \label{rlAgent}
\vspace{-5pt}
\end{figure}

Recently, RL has been shown to be capable of systematically choosing action sequences to conduct AMG \cite{anderson2018a}, \cite{ebrahimi2021a}, \cite{fang2019a}. Figure \ref{rlAgent} provides a conceptual overview of the RL at time \textit{t}, where the agent applies a set of actions $A_t$ to the environment generating the state $S_t$ and reward $R_t$ pair which are then used by the agent to determine the next action or set of actions to apply, or whether the optimal configuration has been reached. The agent-environment interaction is composed of the following components:

\begin{itemize}
    \item \textbf{Agent:} Selects actions that maximize future rewards.
    \item \textbf{Environment:} Generates the states and reward based on the actions taken by the agent at each step.
    \item \textbf{State:} Current configuration of the agent in the environment.
    \item \textbf{Actions:} A set of options, or moves, available to the agent.
    \item \textbf{Reward:} A positive, negative, or zero value returned after each action is applied.
\end{itemize}
\vspace{5pt}

In the AMG scenarios, the primary objective is to develop techniques that can robustify malware detectors against adversarial attacks \cite{anderson2018a}, \cite{ebrahimi2021a}, \cite{ling2021a}, \cite{demetrio2020a}. This requires the generation of evasive malware samples that can avoid detection by the detectors. However, attackers need to minimize their interactions with the environment (detector) to avoid detection and possible countermeasures by the defender \cite{rosenberg2019a}. Accurately emulating the defender, for instance, by implementing a query limit, is crucial for the generation of evasive adversarial variants. RL provides an effective approach to operate under such limitations. To this end, many RL  approaches adopt Deep-Q Networks (DQN), which have been shown to be highly efficient and thus require fewer interactions with the environment \cite{anderson2018a}, \cite{ebrahimi2021a}, \cite{fang2019a}.

Deep Reinforcement Learning (DRL) is the combination of RL and deep neural networks (DNN), to learn policies for decision-making in complex scenarios \cite{mnih-a}, \cite{zhou2021a}. Deep Q-Network (DQN), a DRL algorithm, was developed by Mnih et al., 2013, and has shown exceptional success in a wide range of applications including video games \cite{mnih-a}, \cite{hasselt2016a}, malware evasion \cite{anderson2018a}, \cite{ebrahimi2021a}, \cite{fang2019a}, and penetration testing \cite{zhou2021a}. DQN uses a multi-layered deep neural network to estimate the optimal action value, or Q-function $Q(\textit{s},\textit{a})$ for a given state \cite{anderson2018a}, \cite{hasselt2016a}.

Traditional RL methods utilize a tabular Q-learning method, consisting of a lookup table of historical actions, but such methods often experience performance issues with large state/search spaces \cite{zhou2021a}, \cite{alom2019a}. DQN attempts to solve this issue by replacing the lookup table with a DNN, in particular a deep convolutional neural network \cite{mnih2015a}. The implementation of DQN by Mnih et al. outperformed top benchmark RL methods in the Atari domain and performed at a level comparable to professional human gamers \cite{mnih2015a}. DQN has demonstrated exceptional success in other applications including video games, malware detection, and penetration testing \cite{anderson2018a}, \cite{zhou2021a}, \cite{mnih2015a}. In the AMG applications, thanks to its sample efficiency, we expect that DQN is a viable choice to decrease the frequency of interactions with the malware detector, minimizing the risk of countermeasures implemented by the defender \cite{anderson2018a}, \cite{fang2019a}.

\section{Research Gaps and Questions}
Based on our literature review, the following research gaps have been identified. First, most current AMG attack methods focus on small additive or editing actions, not complete obfuscation, potentially limiting the ability to robustify DL-based malware detectors. Second, it is unclear from a methodological perspective if RL can be used in conjunction with obfuscation tools to modify malware resulting in functional adversarial variants. To address the identified gaps, the following research question is proposed:

\begin{figure*}[t!]
\centering
    \vspace{-9pt}
        \includegraphics[width=0.8\textwidth]{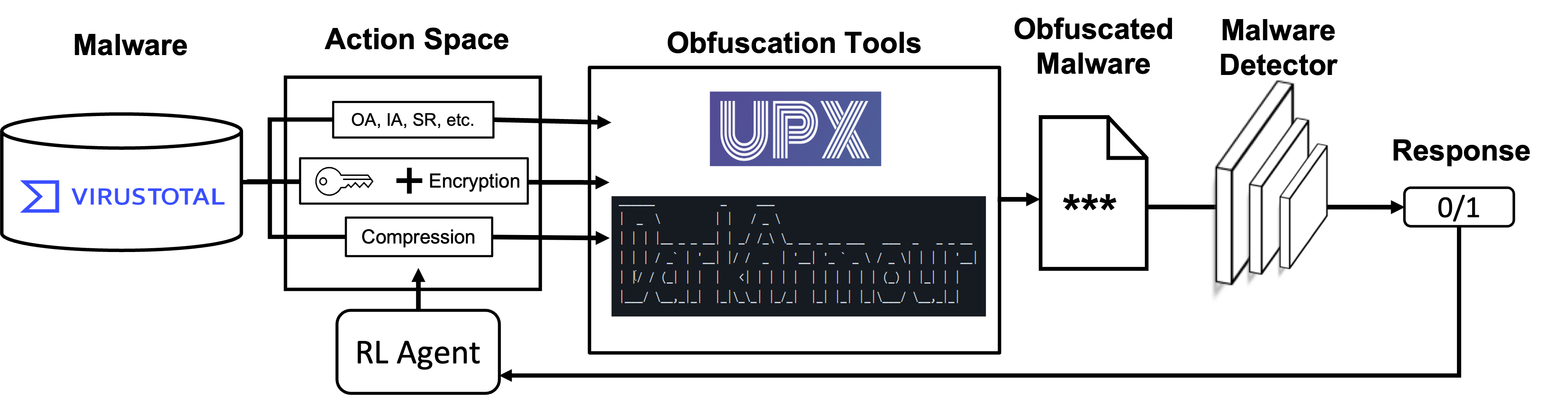}
\vspace{-9pt}
\caption{The OBFU-mal architecture}
\label{Obfumal}
\vspace{-5pt}
\end{figure*}

\begin{itemize}
    \item How can RL, in combination with obfuscation, be used to obfuscate malware and generate evasive functional variants?
\end{itemize}

Motivated by this question, we propose OBFU-mal, a deep RL-based framework to automate the obfuscation of malware for adversarial variant generation.

\section{Proposed Method}
We first introduce the threat model under which OBFU-mal operates in line with previous AMG studies \cite{ebrahimi2021a}, \cite{song2022a}, \cite{carlini2019a}, \cite{hu2021a}. Then, we present our proposed OBFU-mal architecture (Figure~\ref{Obfumal}). Third, we detail the malware testbed and extended action space containing obfuscating actions. Finally, we present the malware detectors used in OBFU-mal’s evaluation.

\subsection{Threat Model}
Following \cite{song2022a, carlini2019a, hu2021a}, our threat model is composed of three main components: adversarial goal, adversarial capability, and adversarial knowledge:

\begin{itemize}
    \item \textbf{Adversarial Goal:} The adversary aims to evade DL-based malware detectors by modifying a known malicious file such that it is recognized as benign.
    \item \textbf{Adversarial Capability:} The adversary is capable of using open-source tools to obfuscate the whole malware file (through encryption). These obfuscation tools can also be utilized by novice hackers with little experience.
    \item \textbf{Adversarial Knowledge:} The adversary operates in a black-box setting \cite{anderson2018a}, in which no knowledge of the malware detection model is assumed. The only available information to the adversary is whether the modified malware was detected or not.
   \end{itemize}

\subsection{The OBFU-mal Architecture}

OBFU-mal's architecture closely emulates the techniques employed by real-world hackers to develop advanced malware. The OpenAI-based gym-malware environment, developed by Anderson et al. \cite{anderson2018a}, serves as the foundation for OBFU-mal, providing a testing ground for generating adversarial samples. Real-world hackers adapt their strategies to evade detection, making this an ideal environment for testing the robustness of malware detectors. OBFU-mal employs an RL agent with a DQN algorithm, mirroring real-world hackers' trial-and-error approach to refining malware. This agent utilizes an extended action space replicating the tactics of malicious actors and incorporates open-source tools for modifying malware samples. This process is illustrated in Figure \ref{Obfumal}, and proceeds as follows. First, the Malware sample is introduced to the RL agent. Then, the RL agent applies actions from the extended action space summarized in Table \ref{actionSpace}. Next, the modified malware sample is assessed by the detector. The feedback from the detector indicates detection or evasion. The process iterates as the next sample is introduced to the agent, which applies actions based on DQN feedback, and the detector evaluates the adversarially generated malware sample.

This approach closely mimics the diverse set of techniques employed by real-world hackers to create obfuscated and adversarial variants. The iterative process, where the detector evaluates the adversarially generated malware samples, reflects the persistent nature of real-world hackers, who continually adjust their techniques based on feedback. As such, OBFU-mal provides a more accurate representation of how real-world hackers try and attack malware detectors. 

\subsection{Action Space}
The extended action space, shown in Table \ref{actionSpace}, details the actions used in select recent AMG studies and includes those added in this study \cite{anderson2018a}, \cite{ebrahimi2021a}, \cite{fang2019a}, \cite{song2022a}.

\begin{table} [ht]
\vspace{3mm}
\centering
\renewcommand{\arraystretch}{1.2}
\setlength{\abovecaptionskip}{0pt}
\setlength{\belowcaptionskip}{-10mm}
\vspace{-6mm}
\caption{Expanded Action Space in OBFU-mal}

\begin{tabular}{
|m{2.3cm}<{\centering}
|m{1.5cm}<{\centering}
|m{3.5cm}<{\centering}
|}

\hline
\textbf{Action} & \textbf{Modification} & \textbf{Description} \\ \hline
Overlay Append & Additive & Append bytes to the end of malware exe \\ \hline
Imports Append & Additive & Add an entry to the import table \\ \hline
Section Rename & Edit & Changes section's name in malware exe \\ \hline
Remove Signature & Edit & Unlink digital signature from certification table \\ \hline
Remove Debug & Additive & Unlink debug section from header \\ \hline
Section Append & Additive & Add a new section to the malware exe \\ \hline
Break Checksum & Edit & Set file's checksum \\ \hline
Change Timestamp & Edit & Change / set timestamp \\ \hline
UPX Pack (Compress) & Obfuscate & Compress malware exe \\ \hline
\textbf{Darkarmour XOR EL1} & \textbf{Obfuscate} & \textbf{Apply one XOR encryption loop} \\ \hline
\textbf{Darkarmour XOR EL2} & \textbf{Obfuscate} & \textbf{Apply two XOR encryption loops} \\ \hline
\textbf{Darkarmour XOR EL3} & \textbf{Obfuscate} & \textbf{Apply three XOR encryption loops} \\ \hline

\end{tabular}
\vspace{2mm}


\label{actionSpace}
\vspace{-3mm}
\end{table}

\begin{table*}[!ht] 
\centering
\renewcommand{\arraystretch}{1.2}
\setlength{\abovecaptionskip}{0pt}
\setlength{\belowcaptionskip}{-10mm}
\caption{State-of-the-Art Benchmark AMG Methods}
\label{benchmarks}
\begin{tabular}{|c|c|c|c|}
\hline

\textbf{Method Category} & \textbf{Method Selected} & \textbf{Description} & \textbf{Reference}\\

\hline

RL & MAB & Append to file and edit sections & Song et al. \cite{song2022a}\\

\hline

RL & AMG-VAC & Append to file and edit sections & Ebrahimi et al. \cite{ebrahimi2021a}\\

\hline

CLM & MalGPT & Appends benign, file-specific perturbations & Hu et al. \cite{hu2021a}\\

\hline

GA & GAMMA & Append to file, file-specific perturbations & Demetrio et al. \cite{demetrio2020a}\\

\hline

{Feature Append} & Benign Append & Appends benign bytes to end of malware file & Castro et al. \cite{castro2019a}\\

\hline

{Feature Append} & Enhanced BFA & Strategically append bytes to file; intent to lower conf. score. & Chen et al. \cite{chen2019adversarial}\\

\hline

{Feature Append} & Random Append & Bytes randomly appended to malware file & Suciu et al. \cite{suciu2019a}\\

\hline

RL & ACER & Append to file and edit sections & Anderson et al. \cite{anderson2018a}\\

\hline

RL & DDQN & Append to file and edit sections & Hasselt et al. \cite{hasselt2016a}\\

\hline

\end{tabular}
\vspace{-5mm}
\end{table*}
\begin{table*}[!hb]
\vspace{-5mm}
\centering
\renewcommand{\arraystretch}{1.2}
\setlength{\abovecaptionskip}{0pt}
\setlength{\belowcaptionskip}{-10mm}
\caption{Evasion Rate Result of OBFU-mal Samples Against MalConv Benchmark AMG Tactics}
\label{malconvEvasion}
\begin{tabular}{|c|c|c|c|c|c|c|c|}
\hline
\multicolumn{1}{|l|}{\textbf{Detector}} & \multicolumn{1}{l|}{\textbf{Method}} & \multicolumn{1}{l|}{\textbf{Botnet}} & \multicolumn{1}{l|}{\textbf{Ransomware}} & \multicolumn{1}{l|}{\textbf{Rootkit}} & \multicolumn{1}{l|}{\textbf{Spyware}} & \multicolumn{1}{l|}{\textbf{Virus}} & \multicolumn{1}{l|}{\textbf{Average}} \\ \hline
\multirow{8}{*}{MalConv}                & ACER                                                               & 37.07\%                              & 25.33\%                                  & 29.41\%                               & 56.09\%                               & 44.76\%                             & 35.04\%                               \\ \cline{2-8} 
                                        & Random Append                                                      & 2.47\%                               & 3.78\%                                   & 4.73\%                                & 2.50\%                                & 1.88\%                              & 2.79\%                                \\ \cline{2-8} 
                                        & BFA                                                                 & 1.14\%                               & 0.11\%                                   & 3.77\%                                & 1.88\%                                & 2.43\%                              & 1.27\%                                \\ \cline{2-8} 
                                        & Enhanced-BFA                                                       & 21.86\%                              & 14.44\%                                  & 3.77\%                                & 11.50\%                               & 12.29\%                             & 15.86\%                               \\ \cline{2-8} 
                                        & GAMMA                                                               & 3.51\%                               & 1.79\%                                   & 0\%                                   & 1.86\%                                & 1.35\%                              & 1.70\%                                \\ \cline{2-8} 
                                        & MalGPT                                                             & 25.86\%                              & 20.33\%                                  & 24.53\%                               & 22.97\%                               & 28.38\%                             & 24.51\%                               \\ \cline{2-8} 
                                        & MAB-malware                                                        & 52.93\%                              & 48.38\%                                  & 45.23\%                               & 50.99\%                               & 55.96\%                             & 50.70\%                               \\ \cline{2-8} 
                                        & \textbf{OBFU-mal (Ours)}                                  & \textbf{69.58\%}                     & \textbf{56.89\%}                         & \textbf{57.80\%}                      & \textbf{66.04\%}                      & \textbf{75.42\%}                    & \textbf{65.15\%}                      \\ \hline
\multirow{6}{*}{LGBM/EMBER}             & DDQN                                                                   & 23.00\%                              & 44.33\%                                  & 39.12\%                               & 19.80\%                               & 27.77\%                             & 28.44\%                               \\ \cline{2-8} 
                                        & ACER                                                                     & 30.99\%                              & 60.11\%                                  & 27.51\%                               & 26.87\%                               & 62.82\%                             & 37.18\%                               \\ \cline{2-8} 
                                        & Enhanced-BFA                                                             & 3.02\%                               & 4.44\%                                   & 4.73\%                                & 5.34\%                                & 6.16\%                              & 3.90\%                                \\ \cline{2-8} 
                                        & AMG-VAC                                                                  & 48.29\%                              & 65.22\%                                  & 61.15\%                               & 29.53\%                               & \textbf{82.40\%}                    & 51.67\%                               \\ \cline{2-8} 
                                        & MAB-malware                                                              & 4.21\%                               & 11.25\%                                  & 4.89\%                                & 10.98\%                               & 24.13\%                             & 11.09\%                               \\ \cline{2-8} 
                                        & \textbf{OBFU-mal (Ours)}                                                 & \textbf{67.32\%}                     & \textbf{92.44\%}                         & \textbf{81.48\%}                      & \textbf{88.07\%}                      & 63.46\%                             & \textbf{79.20\%}                      \\ \hline
\end{tabular}
\end{table*}

The actions used in extant research only impact small portions of the file with the exception of UPX which operates on the whole file. The inclusion of UPX compression, which affects the entire file, mirrors its common use in both legitimate and malicious software. However, malware compressed by UPX tends to be categorized as malicious by most malware detectors \cite{aghakhani2020a}, leading to a decrease in evasion rate. OBFU-mal introduces the actions "Darkarmour XOR EL1/2/3" that utilize one or more loops of eXclusive OR (XOR) encryption, a boolean operator used in cryptography to obfuscate the malware file. The XOR loops are applied by the open-source tool Darkarmour which modifies the malware in a way such that the execution does not require bytes touching the disk \cite{halls2021a}. XOR encryption adds intricacy to the code, making it more challenging for detectors to recognize patterns in the malware.

\section{Evaluation}
\subsection{Experimental Design}
To test the utility and effectiveness of OBFU-mal, we conducted three experiments. First, the evasiveness of OFBU-mal-generated variants was tested against two well-known malware detectors: MalConv and LightGBM (LGBM)/EMBER. Results are compared against a total of nine state-of-the-art benchmark AMG tactics detailed in Table \ref{benchmarks}. Attack method category includes RL, Genetic Algorithm (GA), Causal Language Model (CLM), and feature append; attack methods include Multi-Armed Bandit (MAB) and AMG-Variational Actor Critic (VAC). For RL-based methods, to ensure a fair comparison, a five-query limit was implemented during attacks on the detector. Moreover, in our second experiment, a qualitative analysis of the action sequences that generated evasive variants was performed, and results were contrasted against related work \cite{ebrahimi2021a}. In our third experiment, an ablation analysis was conducted for OBFU-mal to assess the contribution of each component of the framework.

A repository of 3,456 malware samples from VirusTotal was used during the experiment, detailed in Table \ref{malwareTestBed}. These malware samples were modified using OBFU-mal, and evaluated against MalConv and LGBM/EMBER, two well-known malware detectors. MalConv, developed by the Laboratory for Physical Sciences, is a DL-based detector utilizing a deep convolutional neural network (CNN) which takes the first 2M bytes of a sample as it's input \cite{raff2018a}. LGBM/EMBER, developed by Endgame, is a signature-based detector which utilizes a gradient-boosted decision tree (GBDT) and is trained on 2,381 features extracted from the malware binary \cite{anderson2018a}.

\begin{table}[h]
\centering
\renewcommand{\arraystretch}{1.2}
\vspace{-1mm}
\setlength{\abovecaptionskip}{0pt}
\setlength{\belowcaptionskip}{-10mm}
\caption{Composition of Malware Testbed}
\begin{tabular}{
|m{1.7cm}<{\centering}
|m{4.7cm}<{\centering}
|m{0.8cm}<{\centering}
|}

\hline

\textbf{Malware Category} &\textbf{Description} &\textbf{\# of Files}\\

\hline


\hline

\textbf{Botnet} & A network of bots connected through the internet & 526\\

\hline

\textbf{Ransomware} & Encrypts data and files, restricting access and usage until decrypted & 900\\

\hline

\textbf{Rootkit} & Grants admin rights to malware authors & 731\\

\hline

\textbf{Spyware} & Allows malware authors to covertly steal personal data & 640\\

\hline

\textbf{Virus} & Corrupts files on the host system & 659\\

\hline

\textbf{Total} & - & \textbf{3,456}\\

\hline
\end{tabular}
\vspace{-2mm}
\label{malwareTestBed}
\end{table}

\begin{table*}[!ht] 
\centering
\renewcommand{\arraystretch}{1.2}
\setlength{\abovecaptionskip}{0pt}
\setlength{\belowcaptionskip}{-5mm}
\caption{Commonly Occurring Action Sequences Applied by OBFU-mal}
\label{actionSequences}
\begin{tabular}{|c|c|c|}
\hline
\textbf{Method}                          & \textbf{Evasive Action Sequences}                & \textbf{\# of Occurrences} \\ \hline
\multirow{5}{*}{OBFU-mal VS MalConv}      & Change TDS \textbf{→} Overlay Append                      & 94                         \\ \cline{2-3} 
                                         & Overlay Append \textbf{→} Overlay Append                  & 53                         \\ \cline{2-3} 
                                         & Overlay Append \textbf{→} Darkarmour XOR EL2              & 38                         \\ \cline{2-3} 
                                         & Change TDS \textbf{→} Darkarmour XOR EL2                  & 38                         \\ \cline{2-3} 
                                         & Change TDS \textbf{→} Overlay Append \textbf{→} Darkarmour XOR El2 & 30                         \\ \hline
\multirow{5}{*}{OBFU-mal VS LGBM/EMBER} & Overlay Append \textbf{→} Darkarmour XOR EL2              & 106                        \\ \cline{2-3} 
                                         & Change TDS \textbf{→} Darkarmour XOR EL2                  & 69                         \\ \cline{2-3} 
                                         & Change TDS \textbf{→} Overlay Append \textbf{→} Darkarmour XOR El2 & 58                         \\ \cline{2-3} 
                                         & Imports Append \textbf{→} Darkarmour XOR EL1              & 57                         \\ \cline{2-3} 
                                         & Overlay Append \textbf{→} Darkarmour XOR EL3              & 52                         \\ \hline
\end{tabular}
\vspace{-5mm}
\end{table*}
\begin{table*}[!hb] 
\centering
\vspace{-3mm}
\renewcommand{\arraystretch}{1.2}
\setlength{\abovecaptionskip}{0pt}
\setlength{\belowcaptionskip}{-10mm}
\caption{Ablation Study of OBFU-mal Components}
\label{ablation}
\begin{tabular}{|c|c|c|c|c|c|c|}

\hline

\textbf{Method}   & \textbf{Botnet}  & \textbf{Ransomware} & \textbf{Rootkit} & \textbf{Spyware} & \textbf{Virus}   & \textbf{Average} \\ \hline

RL without obfuscation \cite{anderson2018a} & 37.07\% & 25.33\%    & 29.41\% & 56.09\% & 44.76\% & 35.04\% \\ \hline

Obfuscation alone \cite{halls2021a} & 53.61\% & 42.44\%    & 45.96\% & 40.63\% & 56.45\% & 46.5\%  \\ \hline

\textbf{RL with obfuscation (Ours)}    & \textbf{69.58\%} & \textbf{56.89\%}    & \textbf{57.80\%} & \textbf{66.04\%} & \textbf{75.42\%} & \textbf{64.32\%} \\ \hline

\end{tabular}
\vspace{-5mm}
\end{table*}


Performance is measured using evasion rate, which is consistent with prior literature \cite{anderson2018a, ebrahimi2021a, song2022a}. Specifically, evasion rate, $E$, is defined in the following equation:
\[E = M_e/M_t\]
Where $M_e$ denotes the number of samples that evaded detection, and $M_t$ denotes the total number of generated adversarial samples. 

\subsection{Experiment Results}

In the first experiment, the evasiveness of OBFU-mal-generated samples was compared against samples generated by the identified benchmark AMG methods. Table \ref{malconvEvasion} summarizes the results of OBFU-mal against benchmark methods attacking the MalConv and LGBM/EMBER malware detectors.

In table \ref{malconvEvasion} we detail the evasion rates obtained by OBFU-mal generated samples. Against the MalConv detector OBFU-mal achieved an overall evasion rate of 65.15\% which is a percentage-point increase of 14.45\%, 30.11\%, and 40.64\% over MAB-malware, ACER, and MalGPT respectively. Samples interrogated by the LGBM/EMBER detector achieved an overall evasion rate of 79.20\%, a percentage-point increase of 27.53\%, 42.02\%, and 68.11\% against AMG-VAC, Enhanced-BFA, and MAB-malware respectively.

Within the table, we also observed variations in OBFU-mal's performance across different malware types. Notable examples include an increased evasion rate of virus files compared to a decreased evasion rate of ransomware files when attacking MalConv. While outside the scope of this paper, we can offer an educated guess regarding the cause of this performance variance: file complexity. Virus files are typically simple scripts that exploit known vulnerabilities. As such, the binary is easily and massively changed with OBFU-mal's obfuscating actions, rendering the file evasive. Meanwhile, malware such as ransomware contains complex sequences that cannot easily be obfuscated, thus being caught by MalConv's byte-level scrutiny. Interestingly, we find the exact opposite phenomenon when attacking LGBM/EMBER. This could be due to the feature-based nature of LGBM/EMBER, making it more attentive to minute alterations in distinctive attributes of the malware, such as the encryption key of ransomware, while being less sensitive to generic signs of obfuscation used by most software products, signs which may be more represented when obfuscating a simpler virus file. However, these are all conjectures, and further research into why these variations exist is exciting but outside the scope of this paper. Regardless of these variations, OBFU-mal outperformed all benchmark AMG methods across all categories, with the exception of viruses in the case of LGBM/EMBER against the AMG-VAC method, in our research testbed.

In the second experiment, a qualitative analysis of action sequences was conducted to identify sequences that generated the most evasive malware variants. Table \ref{actionSequences} presents the most commonly occurring sequences used in the evasion of MalConv and LGBM/EMBER as applied by OBFU-mal. Obfuscation actions Darkarmour XOR EL1, EL2, and EL3 were observed in the majority of evasive action sequences. This appears to indicate that the newly introduced obfuscation actions are effective in the generation of evasive variants; the aforementioned actions largely appear as the final action in the sequences they were included in.

Lastly, in the third experiment, the impact of including obfuscation in the OBFU-mal framework is assessed through an ablation study. The individual contributions of obfuscation alone \cite{halls2021a}, RL by itself \cite{anderson2018a}, and the combined impact of obfuscation applied alongside the RL actions, developed by Anderson et al. \cite{anderson2018a}, against MalConv were evaluated. Table \ref{ablation} illustrates the contribution of each component assessed in the experiment. RL outperformed obfuscation in only one malware category, spyware; however, the combination of obfuscation and RL outperforms both individual methods. Through the integration of obfuscation into the RL framework enabled by OBFU-mal, we see an average of 17.82\% increase in evasion rates.

Overall, the experiments show that the inclusion of obfuscation actions, as applied by OBFU-mal, significantly improves the evasion of malware against malware detectors, not limited to DL-based, and outperforms benchmark methods. Additionally, the utilization of relatively simple open-source software is sufficient to evade most state-of-the-art detectors, and access to advanced ML- or DL-based methods may not be necessary to create evasive variants.


\section{Case Study}
The majority of extant AMG research is focused on evading open-source malware detectors such as MalConv and LGBM/EMBER, with commercial and pseudo-commercial detectors understudied. One such detector is ClamAV (\url{https://www.clamav.net}), a pseudo-commercial open-source malware detector that also provides endpoint security. The previously documented experiments suggest that attackers can create adversarial variants against open-source DL-based malware detectors, and in turn, those samples are likely to be evasive against ClamAV (i.e., high-quality adversarial variants are transferable). From this, we can test the efficacy of OBFU-mal in a real-world scenario by testing adversarially crafted samples against ClamAV.

\begin{figure}[!ht]
\centering
    \vspace{-2pt}
        \includegraphics[width=0.46\textwidth]{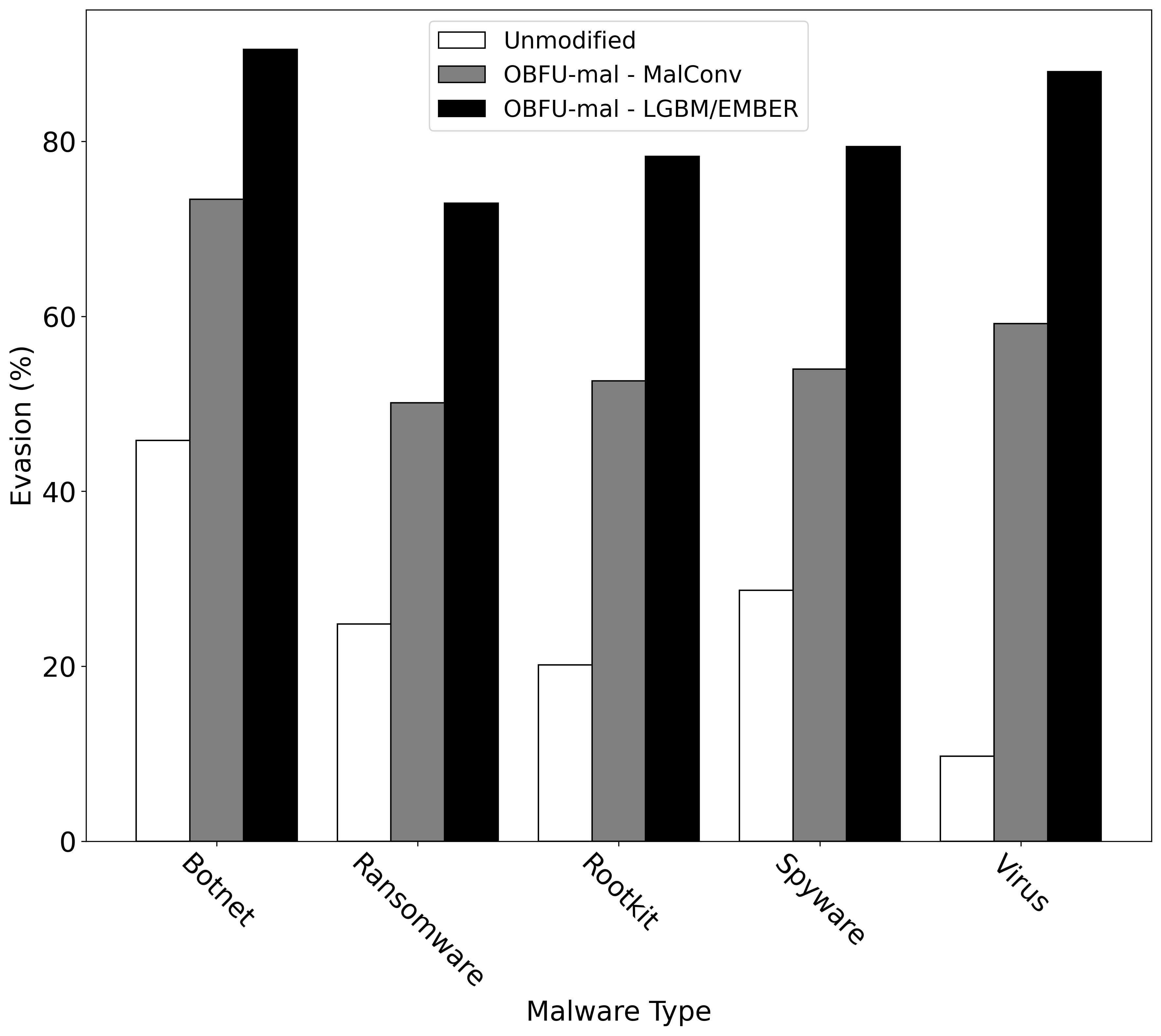}
    \vspace{-5pt}
    \caption{Evasion rates against ClamAV by OBFU-mal}
    \label{clamAvObfu}
\vspace{-3mm}
\end{figure}

Figure \ref{clamAvObfu} highlights OBFU-mal's performance against ClamAV showing evasion rates for unmodified malware samples and malware modified against LGBM/EMBER and MalConv. Samples crafted against LGBM/EMBER and MalConv evade ClamAV 86.44\% and 69.67\% of the time respectively. These results show the transferability of the malware samples generated by OBFU-mal, as samples generated against both LGBM/EMBER and MalConv were still able to evade CLamAV at a higher rate than their unmodified counterparts despite the lack of knowledge of ClamAV during the sample generation process. 

\section{Implications}
Given OBFU-mal's performance against established malware detectors, concerns about its ethics will naturally arise. However, previous adversarial literature has shown that a better understanding of an adversary will yield better defense against such adversaries \cite{li2022assessing}. Given that OBFU-mal seeks to model real-world malware adversaries and their obfuscation methods, we surmise that defenses built upon OBFU-mal's generated adversarial malware samples will be better equipped against real-world hackers. 

We can offer a potential way such a defense may be constructed: adversarial retraining. This refers to finetuning the initial detector (MalConv or LGBM for this study) on evasive adversarial malware samples (i.e., those assigned a benign label by the original detector) and their correct, malicious labels \cite{rathore2021robust}. With this in mind, it becomes clear that adversarially retraining a detector with OBFU-mal samples forces a detector to learn the obfuscating tactics of real-world hackers, thus making them more robust in real-world attacks. While such a defense method could be promising, OBFU-mal is a method focused on offensive AMG and generating evasive adversarial malware samples. Thus, adversarial training and its effect on malware detectors are also outside the scope of this paper. 

\section{Conclusion and Future Work}

AMG studies provide valuable insight into improving anti-malware engines against the application of targeted modifications to malware. However, the vast majority of extant research mostly outlines subtle additive and editing techniques, real world tactics and methods such as obfuscation, encryption and utilization of open source tools are rarely reviewed.

Through the inclusion of obfuscation applied by open-source tools in AMG studies, there may be potential to open many additional avenues for the robustification of DL-based detectors. This work shows that an RL agent can be coupled with open-source obfuscation tools and generate evasive variants capable of evading DL- and signature-based detectors. Automation through RL allows for more efficient exploration of the large action space created when considering open-source tools.

Our proposed OBFU-mal framework, successfully outperformed the vast majority of benchmark methods through obfuscation. This shows that open source tools, requiring little to no advanced training, are sufficient in bypassing state-of-the-art detectors. Moreover, we were able to evade a pseudo-commercial antivirus tool, ClamAV, successfully demonstrating the generalizability of our method.

Promising future directions include using sophisticated source-code obfuscation methods enabled by reverse engineering malware binaries to enhance the AMG capabilities and the adversarial robustness of deep learning-based malware detectors. Another direction is the aforementioned exploration of OBFU-mal's performance variation over certain malware categories, which may yield insight into the strengths and weaknesses of certain malware detectors. Lastly, OBFU-mal's contribution to defensive operations is another exciting yet unexplored direction.

\section*{Acknowledgment}
We would like to sincerely thank VirusTotal for providing the malware dataset and granting access to the corresponding APIs for functionality assessment.

This material is based upon work supported by the National Science Foundation (NSF) Secure and Trustworthy Cyberspace program (grant No. 1936370).

\bibliographystyle{IEEEtran}
\bibliography{main}

\vspace{12pt}

\end{document}